\documentclass[11pt]{article}
\usepackage[margin=1in]{geometry}
\usepackage{graphicx}
\usepackage{subcaption}
\usepackage{amsmath}
\usepackage{multirow}

\def\beq{\begin{equation}}
\def\eeq#1{\label{#1}\end{equation}}
\def\eeqn{\end{equation}}

\def\beqa{\begin{eqnarray}}
\def\eeqa#1{\label{#1}\end{eqnarray}}
\def\eeqan{\end{eqnarray}}

\let\bar=\overbar

\def\Dslash{\not{\hbox{\kern-4pt $D$}}}
\def\dslash{\not{\hbox{\kern-2pt $\del$}}}

\def\msb{{\bar{\ssstyle M \kern -1pt S}}}

\def\Title#1{\begin{center} {\Large {\bf #1} } \end{center}}
\def\Author#1{\begin{center} {\normalsize {\sc #1} } \end{center}}
\def\Institution#1{\begin{center} {\normalsize {\it #1} } \end{center}}
\def\Abstract#1{\noindent {\normalsize {\bf Abstract:} {\normalfont #1}}}
\def\Conference{\vspace{4mm}\begin{raggedright} {\normalsize {\it Talk presented at the 2019 Meeting of the Division of Particles and Fields of the American Physical Society (DPF2019), July 29--August 2, 2019, Northeastern University, Boston, C1907293.} } \end{raggedright}\vspace{4mm}}

\begin{document}

%
%

\Title{Magnetic Field Measurement and Analysis for the Muon g-2 Experiment}

\Author{Ran Hong \footnote{Current institution: University of Kentucky, Department of Physics \& Astronomy, 177 Chem.-Phys. Building, University of Kentucky, 505 Rose Street, Lexington, KY 40506-0055},\footnote{Current email: hongran86@uky.edu} \\on behalf of the Muon g-2 Collaboration \footnote{https://muon-g-2.fnal.gov/collaboration.html}}

\Institution{Argonne National Laboratory, High Energy Physics Division\\
9700 S Cass Ave, IL 60439, USA}

\Abstract{The Muon g-2 Experiment (E989) at Fermilab measures the muon magnetic anomaly, aiming to resolve the greater than 3$\sigma$ discrepancy between the previous measurement and the Standard Model calculation with an improved precision of 140 part-per-billion (ppb). In E989, the muon beam is stored in a ring magnet. The spin precession frequency $\omega_{a}$ is measured by counting the decay positrons in 24 calorimeters, and the magnetic field is measured by nuclear magnet resonance (NMR) probes. An in-vacuum field scanning system consisting of NMR probes and read-out electronics has been implemented to measure the magnetic field applied to the muon beam. An additional 378 NMR probes, placed at fixed locations outside the vacuum chamber, monitor the field drift in between field scans. A high-accuracy probe was designed for calibrating the probes in the scanner. In this presentation, the magnetic field measurement hardware system and analysis methods will be described in detail. The progress of the Run-1 data analysis and improvements in Run-2 will be presented as well.}

\Conference

%
%

\section{Introduction}

\begin{figure}[htb]
\centering
\includegraphics[width=0.7\linewidth]{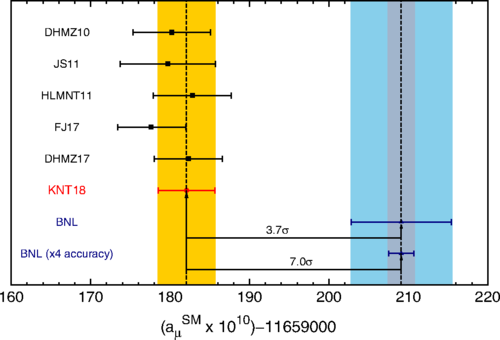}
\caption{Recent theoretical calculations of the muon magnetic anomaly $a_{\mu}$ in the Standard Model.  The result of the experiment E821 and the projected uncertainty of E989 are also shown. This figure comes from Reference~\cite{PhysRevD.97.114025}.}
\label{Fig_TheorySummary}
\end{figure}

The $g$-factor of a lepton in the definition of the magnetic moment
\begin{align}
    \vec{\mu} = g\frac{e}{2m}\vec{S}
\end{align}
equals 2 if no quantum corrections are considered. Due to quantum corrections \cite{PhysRev.73.416}, the $g$-factor deviates from 2, and the lepton magnetic anomaly $a=(g-2)/2$ describes the amount of such deviation. The agreement between the measured electron anomaly $a_{e}$ and the Standard-Model (SM) prediction at the part-per-trillion level has been a benchmark for the validity of quantum electrodynamics (QED) \cite{PhysRevLett.100.120801}. For muons, $a_{\mu}$ was measured and published most recently in 2006 with a precision of 540~ppb at Brookhaven National Lab (Experiment E821) \cite{PhysRevD.73.072003}, and its final result was different from the SM prediction by 2.5 standard deviations \footnote{After 2006, new SM predictions with smaller uncertainties have been developed and the difference becomes larger.}. This difference motivated the physics community to search for mechanisms beyond the SM \cite{PhysRevD.98.055015} that can explain the difference, as well as to perform more accurate calculations of the contributions to $a_{\mu}$ from the known physics effects within the SM \cite{PhysRevD.97.114025,refId0,Davier:2017zfy,davier2019new}. Recently, there has been great progress in calculating the contributions from hadronic vacuum polarization diagrams and hadronic light-by-light diagrams using Lattice-QCD  \cite{Meyer:2018til} and dispersion relation \cite{Colangelo2017}. A few recent results are shown in Figure~\ref{Fig_TheorySummary}, and the difference between the theoretical calculations and the previously measured value persists. The Muon $g-2$ Experiment at Fermilab (E989) \cite{Grange:2015fou} is aiming at to reduce the experimental uncertainty down to 140~ppb, a factor-of-4 improvement from E821. Provided E989 measures the same central value of $a_\mu$ as in E821, the new result will resolve the discrepancy with a greater confidence.
To achieve this goal, E989 will record 21 times the E821 data set, while various systematic uncertainties can be reduced by improved instrumentation. 

In E989, a polarized muon beam is stored in a storage ring with a uniform magnetic field. Because $g\neq2$, the spin precession angular velocity $\vec{\omega}_s$ is different from the cyclotron angular velocity $\vec{\omega}_c$. Assuming that the magnetic field is perfectly uniform and the betatron oscillations of the beam are neglected, the difference of $\vec{\omega}_s$ and $\vec{\omega}_c$ is \cite{PhysRevD.73.072003}
\begin{align}
\vec{\omega}_a=\vec{\omega}_s-\vec{\omega}_c=-\frac{e}{m_{\mu}}\left[a_{\mu}\vec{B}-\left(a_{\mu}-\frac{1}{\gamma^2-1}\right)\frac{\vec{\beta}\times\vec{E}}{c}\right],
\label{Eq_SpinPrecession}
\end{align}
where $\vec{B}$ and $\vec{E}$ are the magnetic and electric fields experienced by the muon, $\vec{\beta}=\vec{v}/c$ is the velocity of the muon relative to the speed of light, and $\gamma=1/\sqrt{1-\beta^2}$. In the muon storage ring, static electric fields are used to vertically focus the muon beam. In order to reduce the magnitude of the $\vec{\beta}\times\vec{E}$ term in Eq.~\ref{Eq_SpinPrecession}, the momentum of the muon beam is chosen to be 3.094~GeV/c ($\gamma=29.3$) so that the coefficient of this term is negligible. In this experiment, the {\em anomalous precession angular frequency} $\omega_{a}$ and the average magnetic field strength $\tilde{B}$ experienced by the muons are measured separately, and the anomalous magnetic moment is proportional to the ratio of the two:
\begin{align}
a_{\mu}=-\frac{m_{\mu}\omega_{a}}{e\tilde{B}}.
\label{Eq_a_mu}
\end{align}
The stored muons ($\mu^{+}$) decay through $\mu^{+}\rightarrow e^{+}\nu_{e}\bar{\nu}_{\mu}$. Since the emitted positrons have less momenta than the muons, they will curve towards the inner side of the ring in the presence of the storage-ring magnetic field. There are 24 calorimeters installed on the inner side of the storage ring, and they measure the arrival time and energy deposition of the incoming positrons. The value of $\omega_{a}$ can be extracted by the counting rate of positrons above a certain energy threshold. The magnetic field in the muon storage region is mapped using proton Nuclear Magnetic Resonance (NMR) probes. The NMR system measures the free-proton precession angular frequency $\omega_{p}$ which is proportional to the magnitude of the magnetic field. The average magnetic field $\tilde{B}$ (or $\tilde{\omega}_{p}$) experienced by the muons is obtained by integrating the $\omega_{p}$ map weighted by the measured muon distribution map. To express $a_{\mu}$ in terms of $\omega_{a}$ and $\tilde{\omega}_{p}$, Eq.~\ref{Eq_a_mu} becomes
\begin{align}
a_{\mu}=\frac{g_{e}}{2}\frac{m_{\mu}}{m_{e}}\frac{\mu_{p}}{\mu_{e}}\frac{\omega_{a}}{\tilde{\omega}_{p}},
\end{align}
where $g_{e}$ is the $g$-factor of an electron, $m_{\mu}/m_{e}$ is the muon-to-electron mass ratio, and $\mu_{p}/\mu_{e}$ is the proton-to-electron magnetic moment ratio. These three values are already measured with uncertainties better than 22~ppb in experiments \cite{PhysRevLett.100.120801,PhysRevLett.82.711,PhysRevLett.35.1619,RevModPhys.88.035009}. 

Two physics runs were completed in 2018 and 2019. In this paper, the structure of the storage ring magnet is described in Sec.~\ref{sec_ring}, and the magnetic field measurement systems are described in Sec.~\ref{sec_measurement_system}. The progress of the $\tilde{\omega}_{p}$ analysis for Physics Run 1, the projected systematic uncertainties, and the upgrades for Physics Run 2 are presented in Sec.~\ref{Sec_Analysis}.

\section{The Storage Ring Magnet}
\label{sec_ring}

\begin{figure}[htb]
\centering
\includegraphics[width=0.9\linewidth]{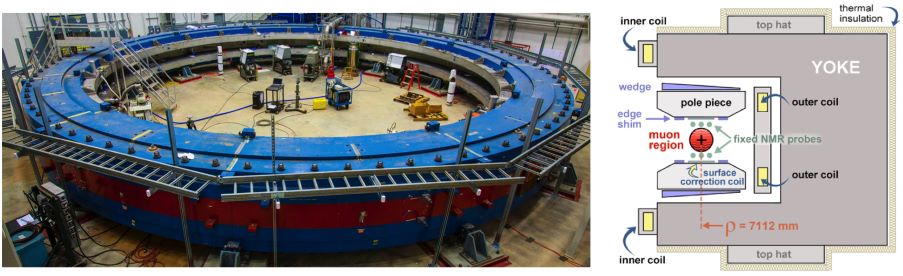}
\caption{The photo of the storage ring magnet and its cross-sectional view. The circular region between the poles represents the muon storage region. The magic radius of $\rho$=7112~mm relative to the center of the ring is labeled. The superconducting coils are marked by the rectangles at the inner and outer radii.}
\label{Fig_MagnetCrossSection}
\end{figure}

The superconducting ring magnet provides a 1.45~T uniform magnetic field to store the muons. The same magnet  \cite{DANBY2001151} used in E821 was transported to Fermilab in 2013. The superconducting coils with their cryostats were transported as a whole, while the iron pieces were disassembled and transported separately and then reassembled. The storage ring magnet and its cross-section are shown in Figure~\ref{Fig_MagnetCrossSection}. The superconducting coils generate a uniform magnetic field in between the inner and outer coil sets, and the iron yoke guides the field lines so that the magnetic field is confined. At the muon beam injection position, there is a tunnel in the back of the iron yoke for the muon beam to enter the magnetic field. In the main field region, the magnetic field points vertically upwards in the space between the pole pieces. To achieve a better uncertainty on $\tilde{\omega}_p$, the magnet has to be shimmed to a higher uniformity than it was in E821. In the magnet shimming campaign of E989, each of the adjustable iron pieces (top/bottom hats, pole pieces, edge shim, and wedge shims) was adjusted carefully to reduce the transverse (radial and vertical directions) and longitudinal (azimuthal direction in the ring) gradients of the field. To shim the non-uniformity at even shorter scales than the sizes of these iron pieces, we used customized iron foils to increase the field strength where the field was weak. About 8500 iron foils were cut precisely to the width that was needed to shim the field at their locations and then epoxied to the surface of the pole pieces. After the 10~month shimming effort, the peak-to-peak variation of the field reached $\pm$25~ppm in the azimuthal direction and $\pm$4~ppm in the transverse directions \cite{Matthias2017}. Besides these passive shimming techniques, 200 concentric coils are placed on the surface of the pole pieces with separately programmable currents. Before each production run period, the magnetic field is scanned, and the values of the currents in these coils are set to cancel the remaining azimuthally averaged transverse field non-uniformity. After optimizing the currents in the coils, the peak-to-peak variation of the azimuthally averaged field cross-sectional map was reduced to 2.5~ppm. The current in the main magnet coils can be actively adjusted based on the field measurements by the field monitoring NMR probes. This feedback mechanism maintains the field at a constant value over a long period.

\section{Magnetic Field Measurement Systems}
\label{sec_measurement_system}

\subsection{The Field Scanning Trolley}
\label{sec_trolley}

\begin{figure}[htb]
\centering
\includegraphics[width=0.9\linewidth]{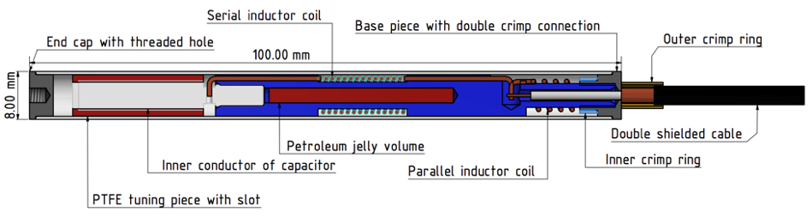}
\caption{Schematics of the trolley probe and the fixed probe.}
\label{Fig_Trolley_Probe}
\end{figure}

\begin{figure}[h]
\centering
  \centering
  \begin{subfigure}[]{0.48\textwidth}
   \includegraphics[width=1\linewidth]{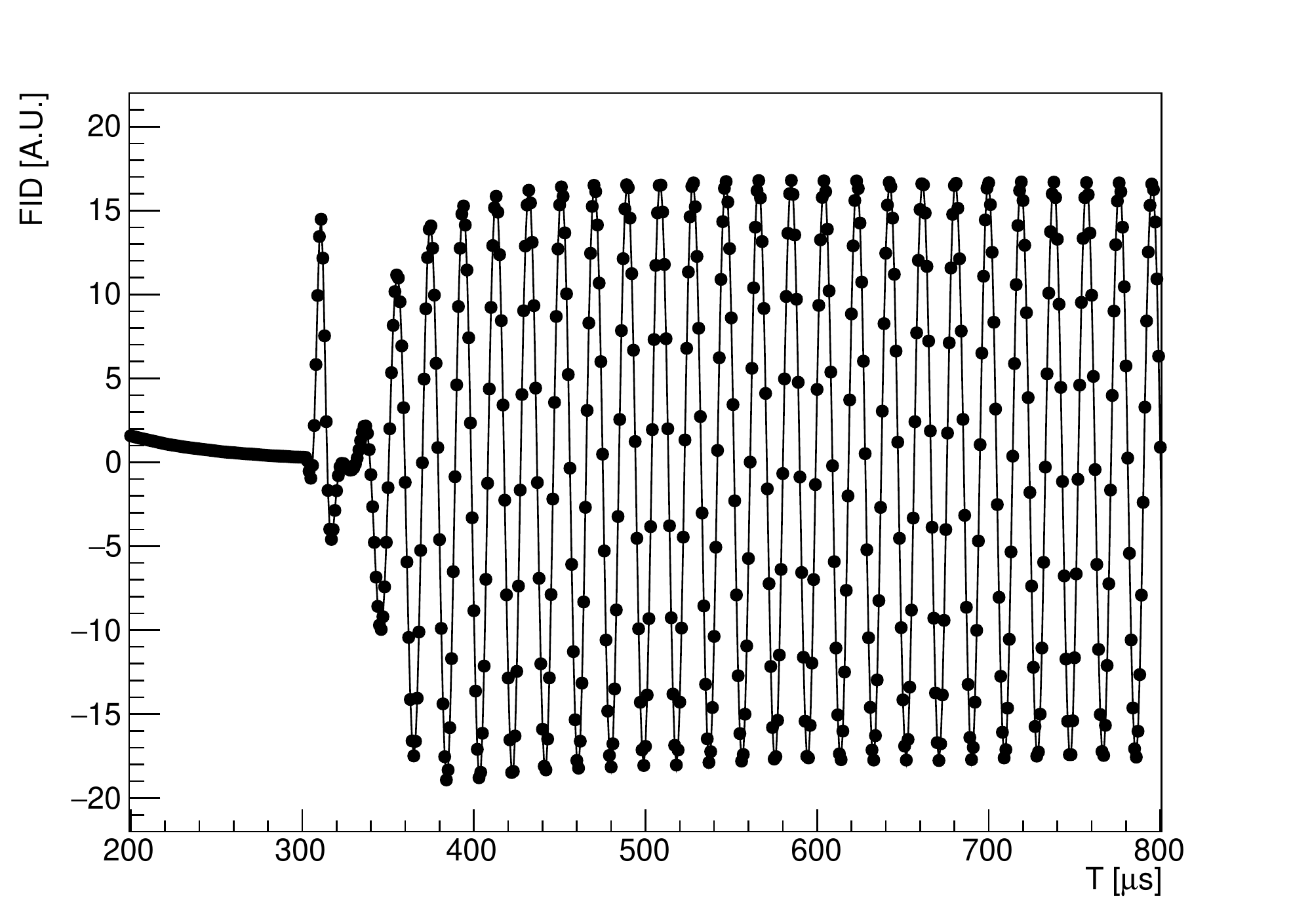}
   \caption{\label{fig:trolley_fid_early}} 
  \end{subfigure}
  \begin{subfigure}[]{0.48\textwidth}
   \includegraphics[width=1\linewidth]{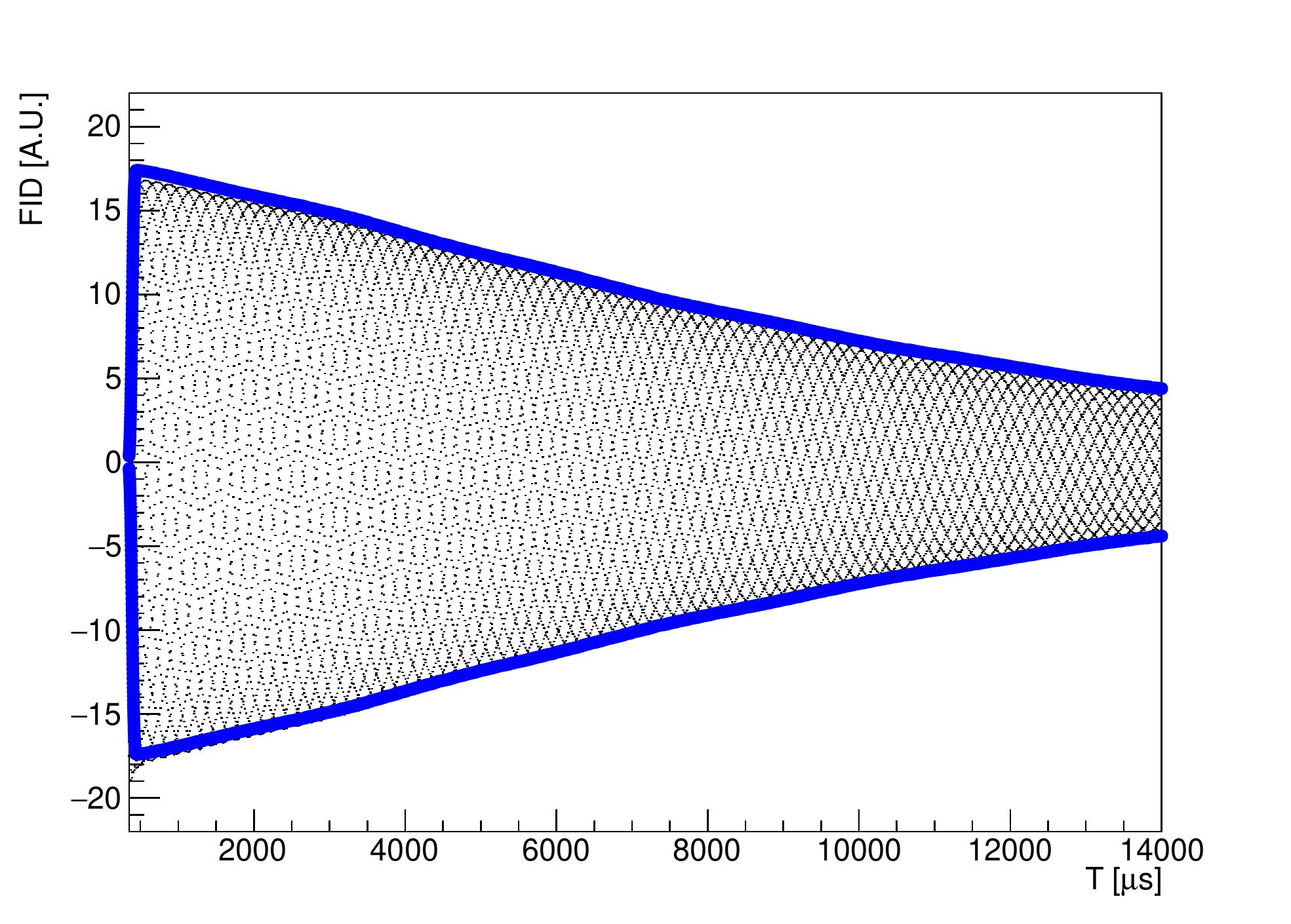}
   \caption{\label{fig:trolley_fid_full}} 
  \end{subfigure}
    \caption{\label{fig:trolley_fid} The digitized FID waveform from the center trolley probe in a magnetic field with extreme uniformity for calibration. The constant pedestal of the ADC is already subtracted. (a) The early section of the waveform showing the start-up features and zoomed-in oscillations. (b) The full waveform from 0.4 to 14~ms, with the envelope highlighted in blue. This Figure comes from Reference~\cite{FID_Paper_2019}.}
\end{figure}

The magnetic field in the muon storage region is scanned in vacuum by a trolley which carries 17 NMR probes and the read-out electronics. The shell of the trolley is the same one\cite{PhysRevD.73.072003} used in E821, but most of the internal components are redesigned to achieve a higher measurement precision. The NMR probes as shown in Fig.~\ref{Fig_Trolley_Probe} are designed and manufactured by the University of Washington. Petroleum jelly is used as the detection material. The new probe holder and read-out electronics were designed by Argonne National Laboratory. The new 3D-printed probe holder fixes the relative position between the probes and the read-out electronics so that magnetic field perturbation generated by the electronics is static. The free induction decay (FID) signals from the NMR probes are fully digitized at a 1~MHz sampling rate and stored for online and offline analyses. The digitized waveform of a typical FID of the trolley probe is shown in Fig.~\ref{fig:trolley_fid}. In the offline analysis, more sophisticated FID frequency extraction algorithms based on the Hilbert Transform are applied, and the systematic uncertainties are also better understood because of the availability of the digitized waveforms. The trolley electronics receive power and communicate with the air-side electronics through a single coaxial cable. The digitization, communication, and control of the trolley electronics are handled by the main NMR board with a SmartFusion 2 system-on-chip \cite{smartfusion2}. This newly designed board can handle much higher data rate than the old one used in E821 while introducing less phase noise in the reference clock for the NMR measurement. The trolley is filled with dry nitrogen to conduct the heat generated by the electronics to the enclosure. Because the entire trolley system is operated in vacuum and therefore the heat is dissipated only through radiation, the total power of the trolley is kept below 1.4~W.

\begin{figure}[htb]
\centering
\begin{subfigure}[]{0.35\textwidth}
   \includegraphics[width=1\linewidth]{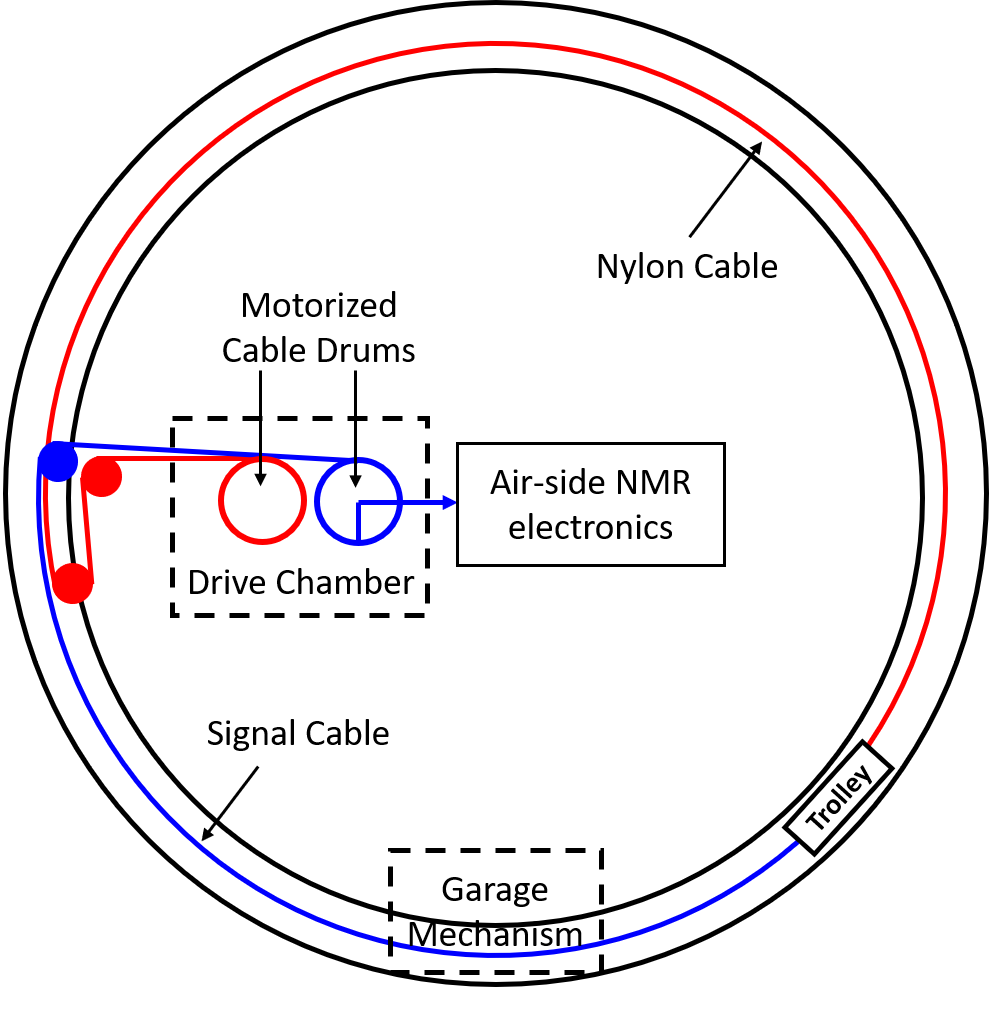}
   \caption{\label{fig:trolley_dirve}Trolley Drive System} 
  \end{subfigure}
  \begin{subfigure}[]{0.6\textwidth}
   \includegraphics[width=1\linewidth]{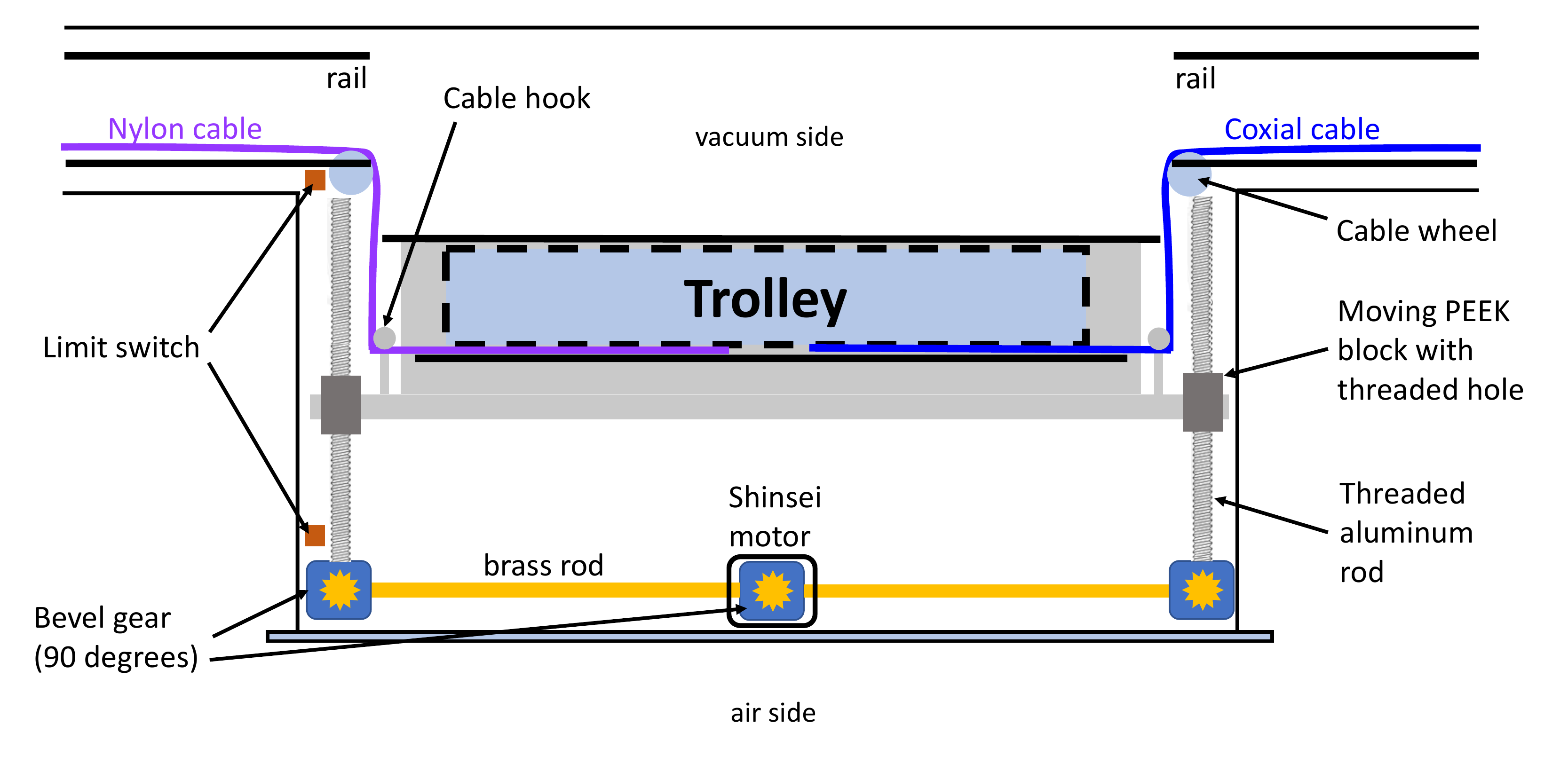}
   \caption{\label{fig:trolley_garage}Garage Mechanism} 
  \end{subfigure}
\caption{Schematics of the trolley drive system and the garage mechanism. This Figure comes from Reference.~\cite{Trolley_Paper_2019}.}
\label{Fig_Trolley_Drive_Garage}
\end{figure}

The field scanning trolley moves inside the storage ring vacuum chamber on rails. Because the rails determine the radial and vertical coordinates of the trolley, the positions and shapes of rails in each chamber section have been aligned at the sub-millimeter level. The motion of the trolley is controlled by two cables pulling on each side. One cable is the coaxial communication cable, and the other is a heavy-duty nylon cable. The cables are guided by three pulleys towards the inner side of the storage ring and strung into a vacuum chamber that contains the cable drums where the cables are wound on. The cable drums are motorized by two remotely controlled motors so that the trolley motion can be programmed and automated. The trolley moves on the rails when one drum is pulling the driving cable and the other drum is releasing cable. At one location of the ring, the trolley can be moved towards the inner side of the ring through a so-called ``garage mechanism'' to avoid the paths of muons and the positrons during the $\omega_{a}$ measurement. The motors that power the cable drums and the garage mechanism are Shinsei piezoelectric motors \cite{shinsei}, which are designed to operate in strong magnetic fields while producing small magnetic footprints. The schematics of the trolley drive system and the garage mechanism is shown in Fig.~\ref{Fig_Trolley_Drive_Garage}. The trolley motion mechanism is essentially the same as the one used in E821, but the motion control system is redesigned to fully automate the motions of the trolley and the garage. The Galil motion control system \cite{galil} was chosen to control the motion of the motors for both the trolley cable drums and the garage. There are rotational encoders reading the position of the cable drums and the garage, and tension sensors reading the tension on the cables. The Galil system uses the encoder and tension values to regulate the motion of the trolley and the garage. The Galil system communicates with a front-end computer of the data acquisition system, and the motions of the trolley and the garage can be controlled remotely via the internet. 

The position of the trolley was determined through the readings of the encoders of the cable drums in E821. However, due to the stretching of the cables, the accuracy and repeatability of this position determination scheme were limited at the scale of $\sim$1~cm. In E989, a new position determination scheme is implemented. Under the inner trolley rail, two rows of bar-codes are printed on the floor. The first row has 2~mm wide black marks evenly spaced, and the second row has a unique pattern for each group. On the bottom of the trolley, a bar-code scanner is installed to detect the reflections of infrared light shone on the bar-code. While the trolley is moving, the bar-code patterns are scanned and recorded together with the NMR waveforms. The bar-code patterns are analyzed offline and the position of each NMR read-out event is determined. Because the bar-codes are stationary relative to the vacuum chamber, the position determined using this scheme is highly repeatable so that the field drift between scans can be determined more reliably. The uncertainty of the position determination of each event is also less than 1~mm.

The muon beam is turned off during the magnetic field scanning, and the entire scan takes $\sim$3 hours. During the stable run period, two field scans were scheduled per week.

\subsection{The Fixed Probes for Monitoring the Field}
\label{sec_fixed_probes}

\begin{figure}[htb]
\centering
\includegraphics[width=0.7\linewidth]{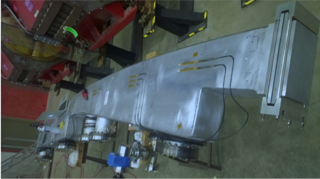}
\caption{Vacuum chambers with monitoring NMR probes installed in the grooves.}
\label{Fig_FixedProbe}
\end{figure}

There are 378 NMR probes installed at fixed locations around the ring to monitor the drift of the magnetic field \cite{Matthias2017}. They are installed in grooves on the vacuum chambers (see Fig.~\ref{Fig_FixedProbe}), above and below the muon storage region as indicated in Fig.~\ref{Fig_MagnetCrossSection}. These probes are identical to the NMR probes in the field scanning trolley. New read-out electronics were designed and manufactured by the University of Washington. These fixed probes measure the magnetic field in the vicinity of the muon storage region during both the field scanning periods and the $\omega_{a}$ measurement periods with the muon beam. In the offline analysis, their readings are used to predict the field drift in the muon storage region. The FID signals from the fixed probes are digitized at a 10~MHz sampling rate and a 4~ms long waveform is recorded for each probe in a read-out cycle. An online analysis function is integrated with the read-out routine on the front-end computer, and the frequencies of all FIDs are extracted. The average of ~$\sim$30 selected probes is used as an estimation of the azimuthally averaged field for the power-supply feedback mechanism. The high-accuracy FID frequency extraction algorithm based on the Hilbert Transform is implemented to the fixed probe FIDs in both online and offline analyses. This algorithm is accelerated using GPU, so that it takes $\sim$0.6~s to process the FIDs of all 378 probes. In the Physics Run 1, the FID read-out and the online analysis are done in series, and the whole sequence takes $\sim$1.67~s, while it was $\sim$10~s in E821. The fast online analysis enables the power-supply feedback mechanism to handle fast field changes better. The finer sampling in time also improves the precision of the field tracking in the offline analysis. 

\subsection{The Absolute Calibration Probe}
\label{sec_plunging_probe}

The magnetic field read by the NMR probes carried by the trolley is perturbed by the material of the trolley shell and electronics, and the material of the probe itself is also magnetized in the field. Because the protons in the NMR probes are in molecules, their NMR frequency is proportional to but different from the free proton NMR frequency. To correct for these effects, we built a cylindrical probe \cite{Flay_2019} with pure water as the detection material. This probe has a well-known magnetic susceptibility chemical shift \cite{RevModPhys.77.1}. The probe is made of materials with a low magnetic footprint, and the perturbation of the probe itself has been measured. This probe is installed on a 3D-motion stage in the storage ring vacuum chamber. It can be plunged into the muon storage region in vacuum to calibrate each NMR probe in the trolley. Thus, this calibration probe is usually referred to as the ``plunging probe''. The plunging probe and its read-out electronics are designed and manufactured by the University of Massachusetts.

\begin{figure}[htb]
\centering
\includegraphics[width=0.9\linewidth]{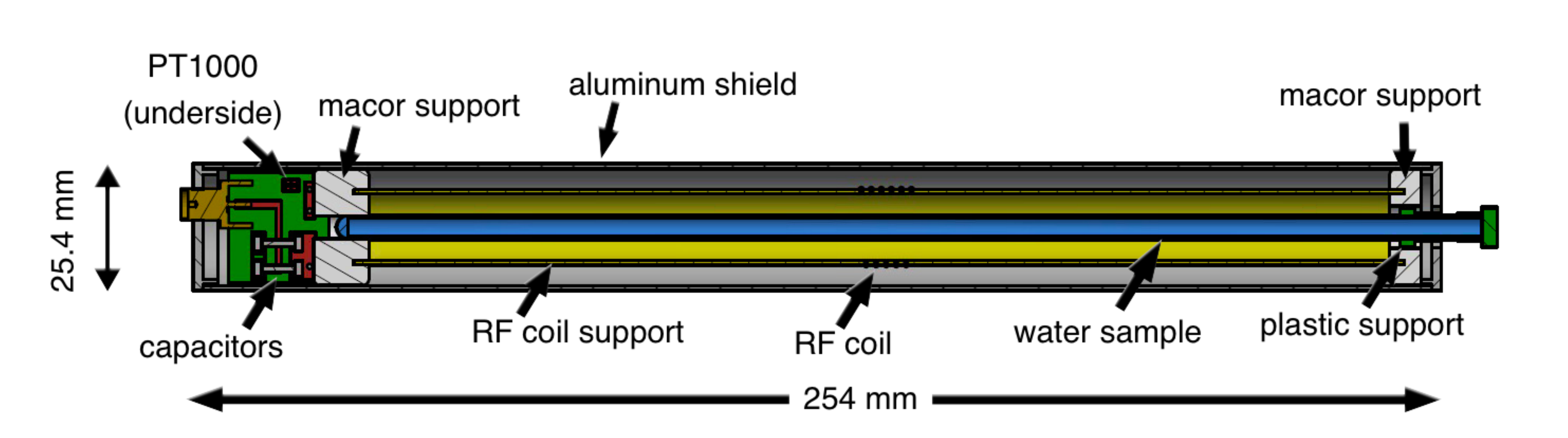}
\caption{Schematics of the plunging probe. This Figure comes from Reference~\cite{Flay_2019}.}
\label{Fig_Plunging_Probe}
\end{figure}

Dedicated calibration runs were conducted at the end of the Physics Run 1. During the calibration, the magnetic field in the calibration region is actively shimmed using the concentric surface correction coils, and the field gradient in the azimuthal direction is shimmed using four special coils in the calibration region. After shimming the field, the trolley and the plunging probe are brought to the calibration position sequentially to read the magnetic field at the same location, and this procedure is repeated for at least four times for each trolley probe. The calibration offset, the difference between the trolley probe FID frequency and the plunging probe FID frequency, is determined for each trolley NMR probe. The calibration offsets and other corrections for the plunging probe are applied to the measured field maps in the field scans, and therefore the free proton precession frequency at each measurement point is determined. 

Besides the plunging probe, a few other calibration probes with spherical shape were built, using water or $^{3}$He \cite{He3Paper_2019,MidhatThesis2019} as detection material. These calibration probes will be cross-calibrated in air for consistency check.

\section{Physics Run 1 Analysis Progress and Improvements for Physics Run 2}
\label{Sec_Analysis}

\begin{figure}[h!tb]
\centering
\includegraphics[width=0.9\linewidth]{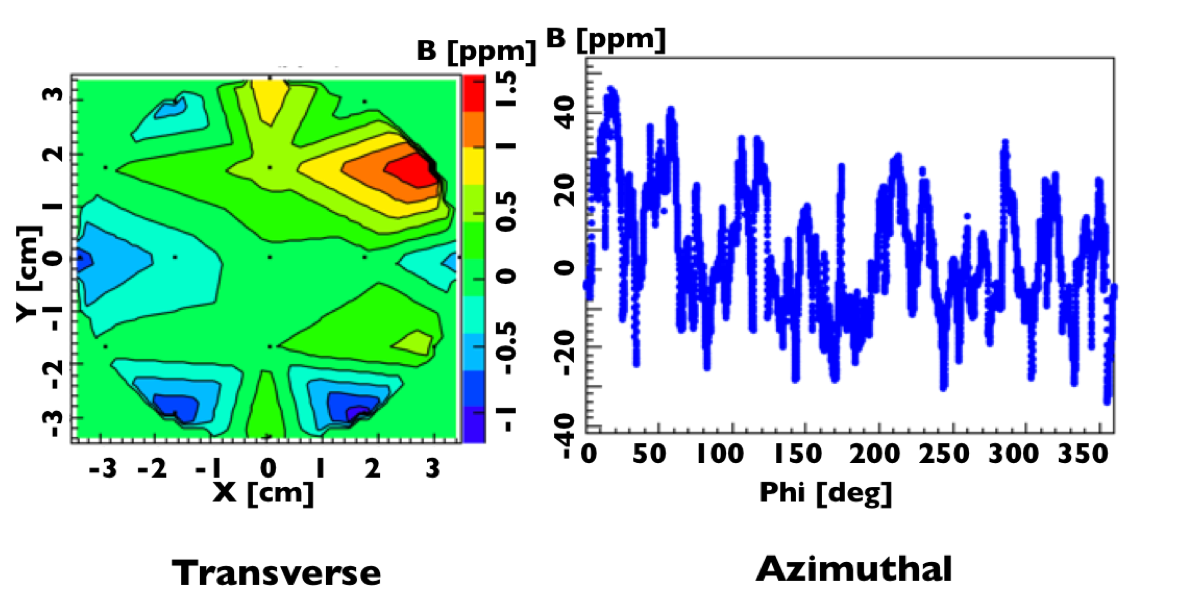}
\caption{(PRELIMINARY) Magnetic field map scanned on May 16th, 2018. The transverse magnetic field distribution is averaged over the azimuthal direction, while each data point in the azimuthal magnetic field distribution is averaged in the transverse direction. Calibration, drift correction and quality controls are not implemented.}
\label{Fig_Omega_p}
\end{figure}

After the rough shimming effort was completed in 2016, the magnetic field measurement systems were installed in early 2017. During the test run from May 2017 to July 2017, the first magnetic field scan was taken. Critical hardware developments and upgrades were identified in this run period and then implemented from August to December in 2017. The commissioning run extended to March 2018, and during this period the FID online analysis, trolley motion control, and the power supply feedback software programs were optimized for user operation and safety. Physics Run 1 started in April 2018 with the magnetic field measurement system operating in a stable condition. By the end of Physics Run 1 in July 2018, more than 20 magnetic field scans were successfully conducted. This includes both regularly scheduled scans and special systematic tests. The magnetic field map from a typical scan is shown in Fig.~\ref{Fig_Omega_p}. 

Offline data analysis software utilities were built using the Art \cite{Green:2012gv} and ROOT \cite{ROOTFramework} data analysis frameworks. The Art-based data production program converts the raw data in the output files from the data acquisition system into ROOT trees and implements the FID frequency extractions and position determinations for the trolley and the plunging probe. Each NMR measurement event is associated with a GPS or Linux system time stamp so that time correlations of these measurements could be constructed in the downstream analysis. The data production program also performs a preliminary data quality screening, tagging suspicious events of hardware failure, unstable magnetic field, and interference from other measurement systems. A more dedicated data quality screening program is developed to accurately determine the periods when the measured magnetic field is unstable or significant failures occurred in the field control and measurement system. These periods are vetoed in both the $\tilde{\omega_{p}}$ and $\omega_{a}$ analyses. 

\begin{figure}[htb]
\centering
\includegraphics[width=0.9\linewidth]{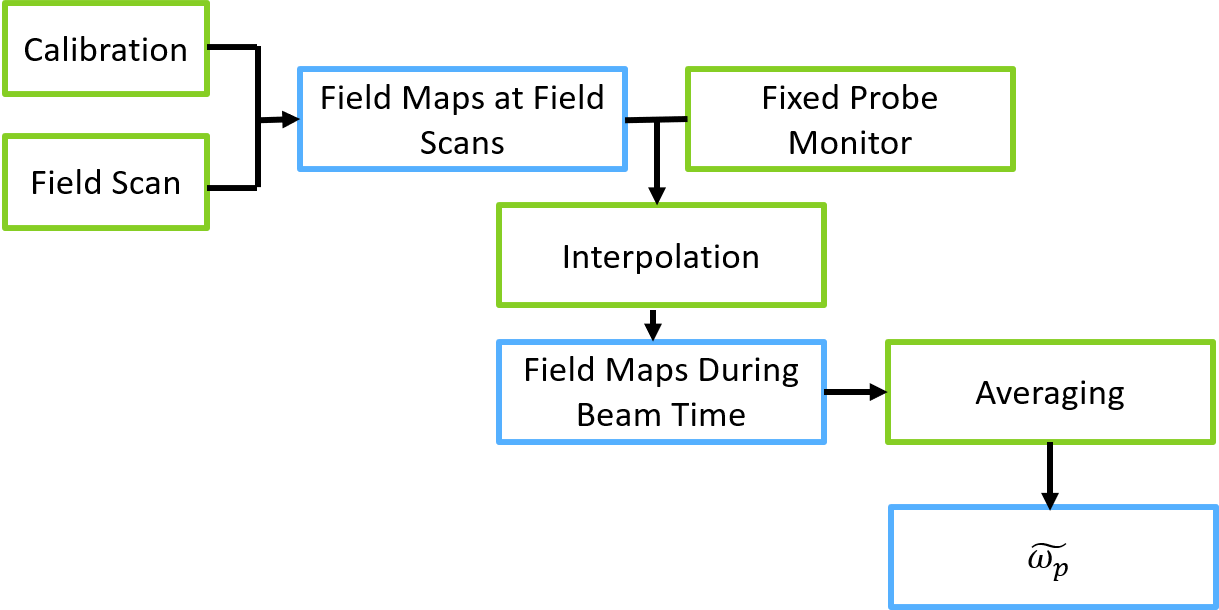}
\caption{Work-flow of the magnetic field data analysis}
\label{Fig_AnalysisFlowChart}
\end{figure}

Taking the produced ROOT trees as the input, the downstream magnetic field analyses (trolley probe calibration, field map construction,
interpolation, and averaging over the muon distribution) were developed by independent analysis teams. The flow-chart of the magnetic field data analysis is shown in Fig.~\ref{Fig_AnalysisFlowChart}. The data analysis for Physics Run 1 is still an on-going effort. The FID frequency extraction, trolley probe calibration, and data quality screening are in advanced stages. The FID frequency extraction algorithm based on Cowan's method \cite{Cowan_1996} is improved and optimized to mitigate the effect of magnetic field inhomogeneity and signal imperfections, such as the time-dependent baselines and distortions. The accuracy of the FID frequency extraction is studied using simulated FIDs signals, and the simulation is designed to model both the response function of the NMR probe and the magnetic field distribution in its sensitive volume. The trolley probe calibration analysis teams are finalizing their systematic uncertainty estimations for magnetic field drift during the calibration and the position alignment between the trolley probe and the plunging probe. The periods with good magnetic field quality have been delivered to the $\omega_{a}$ analysis teams as the input of their data quality control. The trolley position determination using the bar-code data is developed and the improvement in position repeatability is confirmed. The analysis teams are still improving the field averaging algorithm and determining the systematic uncertainties induced by the motion of the trolley. The main challenges in the magnetic field analysis are field map interpolation between field scans and averaging the measured field over the muon beam distribution. Analysis teams have developed their software tools, but more systematic study runs in Physics Run 2 are needed for a more accurate determination of the systematic uncertainties in these analyses. The projected systematic uncertainty budgets are listed in Table.~\ref{tab_systematics}, and the full analysis of Physics Run 1 is expected to be completed at the end of 2019 or beginning of 2020.

\begin{table}[h!]
    \centering
    \begin{tabular}{|p{0.3\linewidth}|p{0.06\linewidth}|p{0.06\linewidth}|p{0.45\linewidth}|}
    \hline
Category                                                                                                                        & E821 (ppb) & E989 (ppb) & Methods                                                                                                   \\
\hline
Absolute probe calibration                                                                                                      & 50         & 35         & More uniform field for calibration                                                                        \\
\hline
Trolley probe calibration                                                                                                       & 90         & 30         & Better alignment between trolley and the plunging probe                                                   \\
\hline
Trolley measurement                                                                                                             & 50         & 30         & More uniform field, less position uncertainty                                                             \\
\hline
Fixed probe interpolation                                                                                                       & 70         & 30         & More stable temperature                                                                                   \\
\hline
Muon distribution                                                                                                               & 30         & 10         & More uniform field, better understanding of muon distribution \\
\hline
Time dependent external magnetic field                                                                                          & -          & 5          & Direct measurement of external field, active feedback                                                     \\
\hline
Higher multipoles, trolley temperature, kicker eddy currents, etc. & 100        & 30         & \begin{tabular}[c]{@{}c@{}}More uniform field, trolley temperature\\   monitor, etc.\end{tabular}          \\
\hline
Total                                                                                                                           & 170        & 70         &     \\   
\hline
\end{tabular}
    \caption{Projected systematic uncertainty budgets for $\tilde{\omega_{p}}$ as stated in the Technical Design Report (TDR) \cite{Grange:2015fou}.}
    \label{tab_systematics}
\end{table}

The Physics Run 2 started in early 2019. Fiberglass insulation was installed onto the magnet to achieve higher magnetic field stability. The major measurement hardware upgrade in Physics Run 2 was the fixed probe trigger module. In Physics Run 1, the fixed probe measurements were triggered asynchronously with the muon beam. The new trigger module reads the trigger pulse from the accelerator complex and then schedules the NMR measurements with programmable delay times. Therefore, a systematic study of the magnetic field measured at different delays relative to the muon fill can be done using this trigger mode. The data acquisition frontend program for the fixed probe system was also upgraded. The measurement and the online analysis were in parallel threads and the time of a measurement cycle was reduced to 1.1~s, shorter than the accelerator cycle (1.44~s). Therefore, all fixed probes can be read out within one accelerator cycle in the synchronized trigger mode. Moreover, during the Physics Run 2 periods, more field scans were performed for systematic studies. For example, consecutive scans were performed within a day, and such scans will improve the determination of the systematic uncertainties of the field map interpolation. The normal field scans were also scheduled at different times of a day so that the field in different hall temperatures were scanned. These systematic studies and improvements in measurements contribute to both the Physics Run 1 analysis and the future Physics Run 2 analysis. 

\section{Conclusion}

The magnetic field measurement systems for the Muon $g-2$ experiment E989 were developed, commissioned, and operated in the Physics Run 1 (2018). Significant improvements in the operations, NMR measurements and position measurement of the field scanning trolley were confirmed. The data analysis of the Physics Run 1 is still on-going. The data production framework has been developed and analysis tasks like the FID frequency extraction and trolley probe calibration are nearly finalized. The magnetic field stability has been improved, and synchronization with the muon beams and NMR measurements has been implemented in the Physics Run 2 (2019). The magnetic field interpolation and the averaging over the muon beam distribution will be improved after including more data focused on systematic studies from Physics Run 2. The analysis of the Physics Run 1 data set is expected to be completed in early 2020.

\section{Acknowledgements}

This work was supported in part by the US DOE, Fermilab and Argonne National Laboratory under contract No. KA2201020. 

\bibliographystyle{ieeetr}
\bibliography{g2bibliography_abbrv}

\begin{thebibliography}{10}

\bibitem{PhysRev.73.416}
J.~Schwinger, ``On quantum-electrodynamics and the magnetic moment of the
  electron,'' {\em Phys. Rev.}, vol.~73, pp.~416--417, Feb 1948.

\bibitem{PhysRevLett.100.120801}
D.~Hanneke, S.~Fogwell, and G.~Gabrielse, ``New measurement of the electron
  magnetic moment and the fine structure constant,'' {\em Phys. Rev. Lett.},
  vol.~100, p.~120801, Mar 2008.

\bibitem{PhysRevD.73.072003}
G.~W. Bennett {\em et~al.}, ``Final report of the {E821} muon anomalous
  magnetic moment measurement at {BNL},'' {\em Phys. Rev. D}, vol.~73,
  p.~072003, Apr 2006.

\bibitem{PhysRevD.98.055015}
P.~Cox, C.~Han, and T.~T. Yanagida, ``Muon $g\ensuremath{-}2$ and dark matter
  in the minimal supersymmetric standard model,'' {\em Phys. Rev. D}, vol.~98,
  p.~055015, Sep 2018.

\bibitem{PhysRevD.97.114025}
A.~Keshavarzi, D.~Nomura, and T.~Teubner, ``Muon $g\ensuremath{-}2$ and
  $\ensuremath{\alpha}({M}_{Z}^{2})$: A new data-based analysis,'' {\em Phys.
  Rev. D}, vol.~97, p.~114025, Jun 2018.

\bibitem{refId0}
{Jegerlehner, Fred}, ``Muon g - 2 theory: The hadronic part,'' {\em EPJ Web
  Conf.}, vol.~166, p.~00022, 2018.

\bibitem{Davier:2017zfy}
M.~Davier, A.~Hoecker, B.~Malaescu, and Z.~Zhang, ``{Reevaluation of the
  hadronic vacuum polarisation contributions to the Standard Model predictions
  of the muon $g-2$ and ${\alpha (m_Z^2)}$ using newest hadronic cross-section
  data},'' {\em Eur. Phys. J.}, vol.~C77, no.~12, p.~827, 2017.

\bibitem{davier2019new}
M.~Davier, A.~Hoecker, B.~Malaescu, and Z.~Zhang, ``A new evaluation of the
  hadronic vacuum polarisation contributions to the muon anomalous magnetic
  moment and to $\mathbf{\boldsymbol\alpha(m_z^2)}$,'' 2019.

\bibitem{Meyer:2018til}
H.~B. Meyer and H.~Wittig, ``{Lattice QCD and the anomalous magnetic moment of
  the muon},'' 2018.

\bibitem{Colangelo2017}
G.~Colangelo, M.~Hoferichter, M.~Procura, and P.~Stoffer, ``Dispersion relation
  for hadronic light-by-light scattering: two-pion contributions,'' {\em
  Journal of High Energy Physics}, vol.~2017, p.~161, Apr 2017.

\bibitem{Grange:2015fou}
J.~Grange {\em et~al.}, ``{Muon (g-2) Technical Design Report},'' 2015.

\bibitem{PhysRevLett.82.711}
W.~Liu {\em et~al.}, ``High precision measurements of the ground state
  hyperfine structure interval of muonium and of the muon magnetic moment,''
  {\em Phys. Rev. Lett.}, vol.~82, pp.~711--714, Jan 1999.

\bibitem{PhysRevLett.35.1619}
W.~D. Phillips, W.~E. Cooke, and D.~Kleppner, ``Magnetic moment of the proton
  in ${\mathrm{h}}_{2}$o in bohr magnetons,'' {\em Phys. Rev. Lett.}, vol.~35,
  pp.~1619--1622, Dec 1975.

\bibitem{RevModPhys.88.035009}
P.~J. Mohr, D.~B. Newell, and B.~N. Taylor, ``Codata recommended values of the
  fundamental physical constants: 2014,'' {\em Rev. Mod. Phys.}, vol.~88,
  p.~035009, Sep 2016.

\bibitem{DANBY2001151}
G.~Danby {\em et~al.}, ``The brookhaven muon storage ring magnet,'' {\em
  Nuclear Instruments and Methods in Physics Research Section A: Accelerators,
  Spectrometers, Detectors and Associated Equipment}, vol.~457, no.~1, pp.~151
  -- 174, 2001.

\bibitem{Matthias2017}
M.~W. Smith, {\em Developing the Precision Magnetic Field for the E989 Muon g-2
  Experiment}.
\newblock PhD thesis, University of Washington, Seattle, 2017.

\bibitem{smartfusion2}
{\em Microsemi/Microchip SmartFusion2 SoC FPGAs}.
\newblock https://www.mouser.com.

\bibitem{FID_Paper_2019}
R.~Hong {\em et~al.}, ``The effects of baseline, signal distortion and noise on
  the fid frequency extraction,'' {\em In preparation}, 2019.

\bibitem{shinsei}
{\em Shinsei Corporation}.
\newblock http://www.shinsei-motor.com/English.

\bibitem{galil}
{\em Galil Motion Control}.
\newblock http://www.galilmc.com.

\bibitem{Trolley_Paper_2019}
P.~Winter {\em et~al.}, ``Design and performance of an in-vacuum, magnetic
  field mapping system for the muon g-2 experiment,'' {\em In preparation},
  2019.

\bibitem{Flay_2019}
D.~Flay and D.~Kawall, ``The high-precision calibration nmr magnetometer for
  the muon $g - 2$ experiment at fermilab,'' {\em In preparation}, 2019.

\bibitem{RevModPhys.77.1}
P.~J. Mohr and B.~N. Taylor, ``Codata recommended values of the fundamental
  physical constants: 2002,'' {\em Rev. Mod. Phys.}, vol.~77, pp.~1--107, Mar
  2005.

\bibitem{He3Paper_2019}
M.~Farooq, T.~Chupp, J.~Grange, A.-T. Booth, D.~Flay, D.~Kawall, and P.~Winter,
  ``Absolute magnetometry with 3he and the muon magnetic moment anomaly
  $g_\mu-2$,'' {\em In preparation}, 2019.

\bibitem{MidhatThesis2019}
M.~Farooq, {\em Absolute Magnetometry with {$^3$He}: Cross Calibration with
  Protons in Water}.
\newblock PhD thesis, University of Michigan, 2019.

\bibitem{Green:2012gv}
C.~Green, J.~Kowalkowski, M.~Paterno, M.~Fischler, L.~Garren, and Q.~Lu, ``{The
  Art Framework},'' {\em J. Phys. Conf. Ser.}, vol.~396, p.~022020, 2012.

\bibitem{ROOTFramework}
{\em ROOT Analysis Framework}.
\newblock https://root.cern.ch/.

\bibitem{Cowan_1996}
B.~Cowan, ``Asymmetric {NMR} lineshapes and precision magnetometry,'' {\em
  Measurement Science and Technology}, vol.~7, pp.~690--695, Apr 1996.

\end{thebibliography}



\end{document}